\documentclass[12pt]{article}
\def\beq{\begin{equation}}
\def\eeq{\end{equation}}
\def\beqa{\begin{eqnarray}}
\def\eeqa{\end{eqnarray}}
\def\ban{\begin{eqnarray*}}
\def\ean{\end{eqnarray*}}
\def\bi{\begin{itemize}}
\def\ei{\end{itemize}}

\begin{document}
\begin{center}
\Large{\bf Applications of the Quon Algebra: 3-D Harmonic
Oscillator and the Rotor Model} \vspace{1.0cm}

{S.S. Avancini, J.R. Marinelli and C. E. de O. Rodrigues} \\
{Depto de F\'{\i}sica - CFM - Universidade Federal de Santa Catarina  \\
Florian\'{o}polis - SC - CP. 476 - CEP 88.040 - 900 - Brazil}
\end{center}
\footnote{E-mail address: ricardo@fsc.ufsc.br}
\begin{abstract}
{In this work we present a method to build in a systematic way a
many-body quon basis state. In particular, we show a closed
expression for a given number N of quons, restricted to the
permutational symmetric subspace, which belongs to the whole
quonic space. The method is applied to two simple problems: the
three-dimensional harmonic oscillator and the rotor model and
compared to previous quantum algebra results. The differences
obtained and possible future applications are also discussed.}
\end{abstract}
\vspace{0.50cm} \noindent PACS number(s): 03.65.Fd: 21.60.Fw \\
keywords: quon algebra, quonic oscillator, rotor model.
\newpage
\vspace{1.0cm}

\section{Introduction}

\bigskip\

Quons are particles that violate statistics by a small amount, which is
controlled by a single parameter q\cite{GreenPRD},\cite{Green6/7/2000}. The
range of variation of the parameter is between -1 and +1, and the limits of
the interval correspond respectively to fermionic an bosonic statistics. The
particular commutation relations obeyed by quons define an algebra (the so
called quon algebra), which, for a single degree of freedom, gives results
very similar to the ones obtained using deformed (or quantum) algebras\cite
{Chaichian}, once we keep the q interval as above defined. For more than a
single degree of freedom however, there are some important differences
between both algebras. One consequence of those differences is that it is
possible to define quonic operators that behave as irreducible su(2) tensors%
\cite{PhysLetA}. In other words, it is possible to assume that the quon
algebra follow the usual angular momentum coupling rules. This last result
opens up the possibility of applying the quon algebra to the study of
many-body systems, with some important technical advantages over deformed
algebras. However, those very same differences also introduce some
complications when we try to build many-body quonic states. The ''q-mutation
'' relation that defines the quon algebra is given by $[a_i,a_j^{\dagger
}]_q=a_ia_j^{\dagger }-qa_j^{\dagger }a_i=\delta _{ij}$ , with the
additional condition $a_i|0>=0$ and $|0>$ being the quon vacuum state.
Unlike quantum algebras, this relation prevent us to establish any
commutation relation between two creation or two annihilation operators,
unless q is either $+1$ or $-1$. Although no such a rule is needed to
calculate vacuum matrix elements of polynomials in the $a^{\prime }s$ or $%
a^{\dagger \prime }s$ \cite{Greenphysica}, the lack of those commutation
relations introduces the necessity to enlarge the basis whenever we try to
define a many-quon state, as we discuss latter in this paper.

It is then our intention here to provide some basis to the use of the quon
algebra as a tool to improve our approximated descriptions of many-body
problems, and at the same time to pose some of the possible differences that
arise with the introduction of that algebra, as compared to the more usually
applied quantum algebras. In fact, the quonic Fock-like Hilbert space
contains all possible symmetry permutations of the polynomials in the $%
a^{\dagger \prime }s$ . As we shall see, even when we restrict ourselves to
the symmetric representation, some important differences between both
algebras can be noticed. Actually, a series of applications using a restrict
quonic subspace was already done in the context of boson mappings\cite{qbos1}%
,\cite{qbos2}. In section 3, we make those differences more explicit here,
through the solution of the three-dimensional harmonic (quon) oscillator as
well as the analysis of the spectrum of a quonic version of the quantum
rotor. Of course, these two examples can be viewed as a simple laboratory to
test our main results in the construction of a many-body quon basis, as
discussed in section 2. We believe that other interesting applications can
be tackled in the future using our present results.

\bigskip\

\section{Many-Body Quon States}

\bigskip\

We start our discussion following the reasonings presented in reference\cite
{Green6/7/2000} for the case of a two quons state. In that case we may write
the following normalized states:

\begin{equation}
\frac 1{\sqrt{1+q}}(a_1^{\dagger })^2|0>,~\frac
1{\sqrt{1+q}}(a_2^{\dagger })^2|0>,~a_1^{\dagger }a_2^{\dagger
}|0> ~\mathrm{{and}~a_2^{\dagger }a_1^{\dagger }|0>.}
\label{2quons}
\end{equation}

The last two states defined in (\ref{2quons}) can be expanded in terms of a
symmetric and a antisymmetric state in the form:

\begin{equation}
a_1^{\dagger }a_2^{\dagger }|0>=\sqrt{\frac{1+q}2}|\phi _s>+\sqrt{\frac{1-q}2%
}|\phi _a>,  \label{2sym}
\end{equation}

\begin{equation}
a_2^{\dagger }a_1^{\dagger }|0>=\sqrt{\frac{1+q}2}|\phi _s>-\sqrt{\frac{1-q}2%
}|\phi _a>,  \label{2anti}
\end{equation}

where,
\[
|\phi _s>=\frac 1{\sqrt{2(1+q)}}(a_1^{\dagger }a_2^{\dagger }+a_2^{\dagger
}a_1^{\dagger })|0~>~~,~~~|\phi _a>=\frac 1{\sqrt{2(1-q)}}(a_1^{\dagger
}a_2^{\dagger }-a_2^{\dagger }a_1^{\dagger })|0~>
\]

We may then conclude that the two-quon basis can be formed by one
antisymmetric and three symmetric states. Also, once any observable must be
represented by a symmetrical operator, the two states in equations (\ref
{2sym}) and (\ref{2anti}) should be considered the same, in the sense that
they give us the same observables. Another way to put this is to recognize
that we can obtain the two-quon (orthonormal) basis by forming the overlap
matrix from the states defined in (\ref{2quons}) and diagonalize it. That
procedure automatically lead us to the three symmetric and one antisymmetric
states above. Then we can diagonalize any observable within the symmetric an
the antisymmetric subspaces separately. Of course, this important property
can be readily generalized to any number of quons. For three quons for
example, besides the well known symmetric and antisymmetric states, there
are more exotic mixed symmetric states. To shorten the corresponding
expressions we adopt the convention $a_i^{\dagger }a_j^{\dagger
}a_k^{\dagger }|0>\equiv |ijk>$ . Then we have for the (normalized) basis
states in that case:

\begin{equation}
|S>=\frac 1{\sqrt{1+2q^2+2q+q^3}}\frac 1{\sqrt{6}%
}[|ijk>+|jik>+|ikj>+|jki>+|kij>+|kji>]  \label{S}
\end{equation}
\begin{equation}
|A>=\frac 1{\sqrt{1+2q^2-2q-q^3}}\frac 1{\sqrt{6}%
}[|ijk>-|jik>-|ikj>+|jki>+|kij>-|kji>]  \label{A}
\end{equation}
\begin{equation}
|MS1>=\frac 1{\sqrt{1-q^2+q-q^3}}\frac 1{\sqrt{12}%
}[|ijk>-|jik>+2|ikj>+|jki>-2|kij>-|kji>]  \label{MS1}
\end{equation}

\begin{equation}
|MS2>=\frac 1{\sqrt{1-q^2+q-q^3}}\frac 12[-|ijk>-|jik>+|jki>+|kji>]
\label{MS2}
\end{equation}

\begin{equation}
|MS3>=\frac 1{\sqrt{1-q^2-q+q^3}}\frac 12[|ijk>-|jik>-|jki>+|kji>]
\label{MS3}
\end{equation}
\begin{equation}
{}|MS4>=\frac 1{\sqrt{1-q^2-q+q^3}}\frac 1{\sqrt{12}%
}[|ijk>+|jik>-2|ikj>+|jki>-2|kij>+|kji>]  \label{MS4}
\end{equation}
\bigskip\

where $i,j,k=1,2,3$ . Also, the cases $i=j$, $i=k$, $j=k$ and $i=j=k$ are
automatically included in expressions (\ref{S}) to (\ref{MS4}) , unless to a
normalization factor which is q-independent. Evidently, the above basis
states can be built from the well known procedure based in the Young tableux
method \cite{Greiner},or, as said before, through the diagonalization of the
overlap matrix obtained from all possible order permutations from the state $%
a_i^{\dagger }$ $a_j^{\dagger }$ $a_k^{\dagger }|0>$ . In fact, the
q-polynomials which appear in the square roots in equations (\ref{S}) to (%
\ref{MS4}), correspond to the eigenvalues of the overlap matrix and measure
the degree of violation of statistics in the system. If we then choose q
sufficiently close to $1(-1)$, we may restrict ourselves to the symmetric
(antisymmetric) subspace, once the observables of the theory do not mix
subspaces corresponding to different symmetries. At this respect it would be
interesting to generalize our above expressions for the symmetric part of
the quonic space. This has a two-fold motivation: first of all many
applications of the deformed algebras (for which only symmetric states are
considered) to many-body problems are restricted to small deformations of
the usual Lie algebra, i.e., q close to 1\cite{BonnaRev} . We would like to
compare some of those results with the equivalent solutions using the quon
algebra. Secondly, the value of q very close ( but not equal) to 1 has the
quite appealing idea to try to take in to account possible violations of
bosonic statistics for systems in which the degrees of freedom are related
to particles with a integer spin value but that are in fact composed by
''fundamental '' fermions.

It is then possible to show (see Appendix) that the most general symmetric
state for a system of N quons can be written as:
\begin{equation}
|n_in_jn_k...;S>=\sqrt{\frac{n_i!n_j!n_k!...}{N![N]!}}{\widehat{S}}%
_N(a_i^{\dagger })^{n_i}(a_j^{\dagger })^{n_j}(a_k^{\dagger })^{n_k}...|0>
\label{NSym}
\end{equation}
where ${\widehat{S}}_N$ is a operator that generates all possible
combinations that are symmetric under the permutation of any of the creation
operators( as defined in the Appendix), $n_i+n_j+n_k+...=N$ and \cite{Kibler}%
:
\begin{equation}
\lbrack N]=\frac{1-q^N}{1-q},  \label{caixote}
\end{equation}
with $[N]!=[N][N-1]....[2][1]$ and [0]!=1. Another important result that we
are going to use next and which is also obtained in the Appendix, is the
following:

\begin{equation}
a_i|n_in_jn_k...;S>=\sqrt{\frac{[N]}N}\sqrt{n_i}|n_i-1,n_jn_k......;S>
\label{aNSym}
\end{equation}

This last expression allows us to calculate matrix elements between
symmetric quonic states with any number of quons. In the following section
we discuss two simple examples and compare them to the deformed algebra
results.

\bigskip\

\section{Applications and Results}

\bigskip\

In order to discuss some examples, it is important to recall that, according
to reference\cite{PhysLetA}, given a set of operators $a_m,a_m^{\dagger }$
for which $m=-j,...,+j$ , and such that they obey the quon commutation
relations, it can be proved that each $a_m^{\dagger }$ behave as a su(2)
irreducible tensor. In other words they obey the expected commutation
relations with angular momentum operators, built from the corresponding
number operators. However, those number operators present a complicated
structure, and are written as an infinite series of the quonic creation and
annihilation operators\cite{GreenPRD}. Also, it is not difficult to obtain a
su(2) scalar from the quons. For example, we may define a quonic three
dimensional harmonic oscillator by the Hamiltonian:

\begin{equation}
H_{osc}^q=\frac{\hbar \omega }2\{(a_{+}^{\dagger }{}a_{+}+a_{-}^{\dagger
}{}a_{-}+a_0^{\dagger }{}a_0)(1+q)+3\}  \label{qosc}
\end{equation}

where
\begin{equation}
a_{+}=\frac 1{\sqrt{2}}(a_1+ia_2)~,~a_{-}=\frac 1{\sqrt{2}
}(a_1-ia_2)~,~a_0=a_3~.
\end{equation}
\noindent  First of all we note that, if $a_1,a_2$ and $a_3$ obey quon
commutation relations, so does the set $a_{+},a_{-}$ and $a_0$ . Secondly,
the factor $(1+q)$ comes from the fact that the Hamiltonian must be
symmetrized and finally we can easily recover the usual harmonic oscillator
Hamilton operator by simply choosing $q=1$, from the above expression. In
order to get the corresponding spectrum we should now diagonalize (\ref{qosc}%
) inside each subspace formed by the states of a given symmetry of the whole
permutation symmetry group and for a given number of quanta $%
N=n_{+}+n_{-}+n_0.$ Alternatively we could proceed with the diagonalization
from the basis formed by all order permutations obtained from the state $%
(a_{+}^{\dagger })^{n_{+}}$ $(a_{-}^{\dagger })^{n_{-}}$ $(a_0^{\dagger
})^{n_0}|0>$. This last procedure amounts however to a diagonalization in a
non-orthonormal basis. On the other hand, the prior diagonalization of the
overlap matrix, as done in the previous section for the $N=3$ case,
corresponds to a partial diagonalization of the Hamiltonian, which then
becomes block diagonal, each block corresponding to a given permutation
symmetry.

As we said before, it is our intention here to make some comparisons to the
deformed algebra results. Then we content ourselves with only the symmetric
part of the solution, which is justifiable once we keep q close enough to 1.
In that case, using equation (\ref{aNSym}) and its Hermitian conjugate, we
readily find for the eigenvalues of our quonic harmonic oscillator, the
remarkable simple result:

\begin{equation}
E_{osc}^q=\frac{\hbar \omega }2\{[N](1+q)+3\}
\end{equation}

Our quonic harmonic oscillator give us then a spectrum which is not equally
spaced but still depends on just one quantum number, the total number $N$.
This is not the case in quantum algebras (see equation (28.38) in reference
\cite{BonnaRev}) , for which the spectrum depends on $N$ and on a additional
quantum number $l$, related to the su$_q$(2) angular momentum. To get an
idea of the effect of the deformation in the spectrum, we present in figure
I a comparison to the regular oscillator for some selected values of the
parameter q.

To better spot the angular momentum structure within the quon algebra we
consider now the quantum rotor. A natural choice in this case is to follow a
Schwinger type of prescription\cite{Sakurai} for the definition of the
angular momentum components:

\begin{equation}
L_{+}=a_{+}^{\dagger }a_{-},~L_{-}=a_{-}^{\dagger
}a_{+},~L_0=\frac 12\{a_{+}^{\dagger }a_{+}-a_{-}^{\dagger
}a_{-}\}  \label{AngDef}
\end{equation}

We again restrict our results to the symmetric quon subspace. Using once
more equation (\ref{aNSym}), we get the results:

\begin{equation}
<n_{+}^{\prime },n_{-}^{\prime };S|[L_{+},L_{-}]|n_{+},n_{-};S>=\frac{[N]}%
N(n_{+}-n_{-})\delta _{n_{+}^{\prime }n_{+}}\delta _{n_{-}^{\prime }n_{-}}
\end{equation}

and,

\begin{equation}
<n_{+}^{\prime },n_{-}^{\prime };S|2L_0|n_{+},n_{-};S>=\frac{[N]}%
N(n_{+}-n_{-})\delta _{n_{+}^{\prime }n_{+}}\delta _{n_{-}^{\prime }n_{-}}
\end{equation}

where now $N=n_{+}+n_{-}$. Using the same type of calculation we may also
prove that:

\begin{equation}
<n_{+}^{\prime },n_{-}^{\prime };S|[L_0,L_{+}]|n_{+},n_{-};S>=<n_{+}^{\prime
},n_{-}^{\prime };S|L_{+}|n_{+},n_{-};S>
\end{equation}

\begin{equation}
<n_{+}^{\prime },n_{-}^{\prime
};S|[L_0,L_{-}]|n_{+},n_{-};S>=-<n_{+}^{\prime },n_{-}^{\prime
};S|L_{-}|n_{+},n_{-};S>
\end{equation}

\noindent The above results show that the operators defined in (\ref{AngDef}%
), behave as genuine angular momentum components within the symmetric
subspace. All we need now is to obtain the matrix element of the operator $%
L^2$, which gives:

\begin{equation}
<n_{+}^{\prime },n_{-}^{\prime };S|L^2|n_{+},n_{-};S>=\frac{[N]}2~(\frac{[N]}%
2+1)~\delta _{n_{+}^{\prime }n_{+}}\delta _{n_{-}^{\prime }n_{-}}
\end{equation}

\noindent Assuming the correspondence $n_{+}=l+m$ and $n_{-}=l-m$ we finally
obtain for our q-rotor spectrum:

\begin{equation}
E_l^q=A\frac{[2l]}2\left\{ \frac{[2l]}2+1\right\}  \label{qrotor}
\end{equation}

with $A$ being the inertia constant. Again, we see that, although we get the
right limit for q=1, the spectrum given by equation (\ref{qrotor}) is
different from the one obtained through quantum algebra techniques (see
equation (19.3) in \cite{BonnaRev}). At this respect, one interesting result
that emerges from the deformed algebra rotor, is its ability to describe
stretching effects as experimentally observed in the spectra of heavy nuclei
and molecules, with the introduction of a single parameter\cite{BonaRotor}.
In our case, we could test the applicability of our previous results doing
the same sort of analysis, using expression (\ref{qrotor}). In figure II we
show the experimental spectrum of the fundamental rotational band in the $%
^{240}$Pu nucleus, chosen here as typical sample, together with the one
obtained from our quonic rotor. We choose a q-value that minimizes the
differences between the theoretical spectrum and the experimental one,
within the interval allowed by the quon algebra. It is interesting to
observe that a q-value slightly smaller than $1$ is enough to produce the
desired modifications in order to take in to account the stretching effects
just mentioned, as we can observe particularly for higher values of the
angular momentum, compared to the rigid rotor result. Also shown in the same
figure is the best fit spectrum obtained using the quantum algebra approach
from reference\cite{BonaRotor}.

\bigskip\

\section{Conclusions}

\bigskip\

In this letter we have discussed and presented a method to build
in a systematic way, a basis of states that represent a system of
identical quons. The novel feature of that type of states lie in
the fact that, once quons obey commutation relations which
interpolate between bosons and fermions, all kind of permutation
symmetries can be accommodated in a many-quon state. The
probability that each type of symmetry occur is then controlled
by a single parameter q. Once any observable should be symmetric
by any particle exchange, a classification of the states by their
permutational symmetry amounts to a partial diagonalization of the
corresponding operator in the whole quonic space. In order to
make some contact with the deformed algebra results, we have kept
our attention to the totally symmetric subspace, for which we
could find a closed general expression for a state with any
number of quons, as well as for the action of an operator on it.
Two simple examples were then considered here within that point
of view: a quonic version of the three-dimensional harmonic
oscillator and a rotor model based on the quon algebra. As a by
product we have found out the interesting result that the angular
momentum operator written in terms of quons and within the
symmetric subspace behaves as usual su(2) angular momentum
operators, having the same functional form as in the case of
regular bosons. Also, a comparison of both examples with the
results previously obtained within quantum algebras, show a very
distinct energy spectrum distribution in the case of the harmonic
oscillator where the same degree of degeneracy observed in the
usual bosonic oscillator is recovered, contrary to what happens
when we use the quantum su$_q$(2) algebra. As for the rigid
rotor, though we have found an energy spectrum very close in both
cases, the angular momentum operator properties are quite
different from the su$_q$(2) properties.

Quantum or q-deformed algebras are by now considered a very powerful tool to
deal with physical systems for which usual algebras can not take in to
account some of their properties. The quon algebra, which is considered in
the literature as a particular case\cite{Kibler}of those algebras, present
subtle differences that may reflect some important and even fundamental
differences in what concerns the interpretation of the final results. As
mentioned in the Introduction, an interesting point that deserves some
investigation in the near future is the analysis of bosonic systems that are
in fact composed by fermions, once some observed deviations from a true
boson behavior could in principle be realized by the quon algebra in a
natural way.

\vskip 0.35in
\begin{center}
\textbf{Acknowledgments}
\end{center}
This work was partially supported by CNPq - Brazil. \vspace{0.5cm}
\begin{center}
\textbf{Appendix}
\end{center}

In this appendix we prove two important results, eqs.(\ref{NSym},\ref{aNSym}%
), given in the main text. We begin with the most general symmetric (not
normalized) N quons state in an arbitrary basis:
\begin{equation}
\widehat{S}_{\mathrm{N}}(a_i^{\dagger })^{n_i}(a_j^{\dagger
})^{n_j}(a_k^{\dagger })^{n_k}...|0>~\equiv ~{\frac 1{{n_i!n_j!n_k!...}}}%
\sum_{P_{\mathrm{N}}}a_{{\alpha }_1}^{\dagger }a_{{\alpha }_2}^{\dagger
}...a_{{\alpha }_{n_i}}^{\dagger }a_{{\alpha }_{n_i+1}}^{\dagger }...a_{{%
\alpha }_{n_i+n_j+1}}^{\dagger }...a_{{\alpha }_N}^{\dagger }|0>~~,
\label{symm}
\end{equation}
where N=$n_i+n_j+n_k+...$ and the summation runs over all the N!
permutations, $P_{\mathrm{N}}$, in the indices $\alpha _1,\alpha
_2,...,\alpha _N$. We order these indices such that $\alpha
_1,\alpha _2,...,\alpha _{n_i}$ corresponds to the i-state,
$\alpha _{n_i+1},\alpha _{n_i+2},...,\alpha _{n_i+n_j}$ to the
j-state and so on. The factorials
under the denominator accounts for repeated terms in the summation and  eq.(%
\ref{symm}) is the definition of the $\widehat{S}_{\mathrm{N}}$ operator. We
now prove by induction the following result:
\begin{equation}
a_i~\widehat{S}_{\mathrm{N}}(a_i^{\dagger })^{n_i}(a_j^{\dagger
})^{n_j}(a_k^{\dagger })^{n_k}...|0>~=~[N]~\widehat{S}_{\mathrm{{N}-1}%
}(a_i^{\dagger })^{n_i-1}(a_j^{\dagger })^{n_j}(a_k^{\dagger })^{n_k}...|0>~,
\label{hypo}
\end{equation}
where [N] is given in eq.(\ref{caixote}). It is easy to show that the
relation above is valid for N=1($\widehat{S}_{\mathrm{0}}=I$, the identity
operator). We assume that it is also true for a N-1 quons state, i. e.,
\begin{equation}
a_i~\widehat{S}_{\mathrm{{N}-1}}(a_i^{\dagger })^{n_i^{\prime
}}(a_j^{\dagger })^{n_j^{\prime }}(a_k^{\dagger })^{n_k^{\prime
}}...|0>~=~[N-1]~\widehat{S}_{\mathrm{{N}-2}}(a_i^{\dagger })^{n_i^{\prime
}-1}(a_j^{\dagger })^{n_j^{\prime }}(a_k^{\dagger })^{n_k^{\prime }}...|0>~,
\label{induc}
\end{equation}
where N-1=$n_i^{\prime }+n_j^{\prime }+n_k^{\prime }+...$ . One can shows
the following property of the symmetrization operator:
\[
\widehat{S}_{\mathrm{N}}(a_i^{\dagger })^{n_i}(a_j^{\dagger
})^{n_j}(a_k^{\dagger })^{n_k}...|0>~=~a_i^{\dagger }\widehat{S}_{\mathrm{{N}%
-1}}(a_i^{\dagger })^{n_i-1}(a_j^{\dagger })^{n_j}(a_k^{\dagger
})^{n_k}...|0>
\]
\begin{equation}
~+~a_j^{\dagger }\widehat{S}_{\mathrm{{N}-1}}(a_i^{\dagger
})^{n_i}(a_j^{\dagger })^{n_j-1}(a_k^{\dagger })^{n_k}...|0>~+a_k^{\dagger }%
\widehat{S}_{\mathrm{{N}-1}}(a_i^{\dagger })^{n_i}(a_j^{\dagger
})^{n_j}(a_k^{\dagger })^{n_k-1}...|0>~+~...~~.  \label{SNpro}
\end{equation}
This property follows from the definition of the symmetrization operator, eq.(%
\ref{symm}), and its rearrangement as given below:
\[
\widehat{S}_{\mathrm{N}}(a_i^{\dagger })^{n_i}(a_j^{\dagger
})^{n_j}(a_k^{\dagger })^{n_k}...|0>~=~{\frac 1{{n_i!n_j!n_k!...}}}
\]
\[
\cdot \big( a_{{\alpha }_1}^{\dagger
}\sum_{P_{\mathrm{{N}-1}}}{\hat{a}}_{{\alpha }_1}^{\dagger
}a_{{\alpha }_2}^{\dagger }...a_{{\alpha }_N}^{\dagger
}|0>~+~a_{{\alpha }_2}^{\dagger }\sum_{P_{\mathrm{{N}-1}}}a_{{\alpha }%
_1}^{\dagger }{\hat{a}}_{{\alpha }_2}^{\dagger }a_{{\alpha }_3}^{\dagger
}...a_{{\alpha }_N}^{\dagger }|0>~
\]
\begin{equation}
~+~...~+~a_{{\alpha }_{n_i}}^{\dagger }\sum_{P_{\mathrm{{N}-1}}}a_{{\alpha }%
_1}^{\dagger }...{\hat{a}}_{{\alpha }_{n_i}}^{\dagger }a_{{\alpha }%
_{n_i+1}}^{\dagger }...a_{{\alpha }_N}^{\dagger }|0>~+~...~+~a_{{\alpha }%
_N}^{\dagger }\sum_{P_{\mathrm{{N}-1}}}a_{{\alpha }_1}^{\dagger }...a_{{%
\alpha }_{N-1}}^{\dagger }{\hat{a}}_{{\alpha }_N}^{\dagger }|0>~\big)~~,
\end{equation}
where the hat symbol on the creation operator means that it is
omitted in that position. The first $n_i$ terms above are equal,
since $\alpha _1,\alpha _2,...,\alpha _{n_i}$ are associated to
the i-state, the same argument may be used for the next $n_j$
terms and so on. So the property given in eq.(\ref{SNpro}) is
proved.

From eq.(\ref{SNpro}) and the q-mutation relation, we obtain for
the action of the annihilation operator, $a_i$, in the N quons
symmetric state the result as follows:
\[
a_i\widehat{S}_{\mathrm{N}} (a_i^{\dagger })^{n_i}(a_j^{\dagger
})^{n_j}(a_k^{\dagger })^{n_k}...|0>~=~\widehat{S}_{\mathrm{{N}-1}}
(a_i^{\dagger })^{n_i-1}(a_j^{\dagger })^{n_j}(a_k^{\dagger })^{n_k}...|0>
\]
\[
~+~q\big(~ a_i^{\dagger} a_i~\widehat{S}_{\mathrm{{N}-1}} (a_i^{\dagger
})^{n_i-1}(a_j^{\dagger })^{n_j}(a_k^{\dagger })^{n_k}...|0>~+~
a_j^{\dagger} a_i~\widehat{S}_{\mathrm{{N}-1}} (a_i^{\dagger
})^{n_i}(a_j^{\dagger })^{n_j-1}(a_k^{\dagger })^{n_k}...|0>
\]
\begin{equation}  \label{aisn}
~+~ a_k^{\dagger} a_i~\widehat{S}_{\mathrm{{N}-1}} (a_i^{\dagger
})^{n_i}(a_j^{\dagger })^{n_j}(a_k^{\dagger })^{n_k-1}...|0> \big) ~
\end{equation}
We now use eq.(\ref{induc}) to rewrite the term between parenthesis and get
\[
a_i \widehat{S}_{\mathrm{N}} (a_i^{\dagger })^{n_i}(a_j^{\dagger
})^{n_j}(a_k^{\dagger })^{n_k}...|0>~=~\widehat{S}_{\mathrm{{N}-1}}
(a_i^{\dagger })^{n_i-1}(a_j^{\dagger })^{n_j}(a_k^{\dagger })^{n_k}...|0>
\]
\[
~+~q[\mathrm{{N}-1]\big(~ a_i^{\dagger} \widehat{S}_{{N}-2} (a_i^{\dagger
})^{n_i-2}(a_j^{\dagger })^{n_j}(a_k^{\dagger })^{n_k}...|0>~+~
a_j^{\dagger} \widehat{S}_{{N}-2} (a_i^{\dagger })^{n_i-1}(a_j^{\dagger
})^{n_j-1}(a_k^{\dagger })^{n_k}...|0> }
\]
\begin{equation}  \label{aif}
~+~ a_k^{\dagger} \widehat{S}_{\mathrm{{N}-2}} (a_i^{\dagger
})^{n_i-1}(a_j^{\dagger })^{n_j}(a_k^{\dagger })^{n_k-1}...|0> \big)~
\end{equation}
Finally using eq.(\ref{SNpro}) in the term between parenthesis and
the q-number property, [N]=1+q[N-1], we may rearrange the above
expression as:
\begin{equation}  \label{eqla}
a_i~\widehat{S}_{\mathrm{N}} (a_i^{\dagger })^{n_i}(a_j^{\dagger
})^{n_j}(a_k^{\dagger })^{n_k}...|0>~=~[N]~\widehat{S}_{\mathrm{{N}-1}}
(a_i^{\dagger })^{n_i-1}(a_j^{\dagger })^{n_j}(a_k^{\dagger
})^{n_k}...|0>~.~~
\end{equation}
and the proof by induction is finished.

Now we obtain the normalized symmetric state. Let us write:
\begin{equation}
|n_i,n_j,n_k,...;S>~\equiv ~A_{n_in_jn_k...}\widehat{S}_{\mathrm{N}%
}(a_i^{\dagger })^{n_i}(a_j^{\dagger })^{n_j}(a_k^{\dagger })^{n_k}...|0>~.
\end{equation}
So the normalization constant,$A_{n_in_jn_k...}$ is determined by:
\[
1=<n_i,n_j,n_k,...;S|n_i,n_j,n_k,...;S>=A_{n_in_jn_k...}^2~{\frac{N!}{{%
n_i!n_j!n_k!}}}
\]
\[
\cdot ~<0|...(a_k)^{n_k}(a_j)^{n_j}(a_i)^{n_i}\widehat{S}_{\mathrm{N}%
}(a_i^{\dagger })^{n_i}(a_j^{\dagger })^{n_j}(a_k^{\dagger
})^{n_k}...|0>~=~A_{n_in_jn_k...}^2~{\frac{{N![N]!}}{{n_i!n_j!n_k!}}}~~,
\]
where in order to get the result above, eq.(\ref{eqla}), was iterated. So
the normalization factor in eq.(\ref{NSym}) is obtained. From eq.(\ref{eqla}%
) and the normalization factor obtained above, eq.(\ref{aNSym}) follows
trivially.

\bigskip\

\newpage

FIGURE\ CAPTIONS

\bigskip\

Figure I. Harmonic oscillator spectrum (\textbf{A}) compared to the quonic
harmonic oscillator spectra obtained for two different values of the
deformation parameter : $q=0.99$ (\textbf{B}) and $q=0.98$ (\textbf{C}).

\bigskip\

Figure II. The experimental (\textbf{A})\  spectrum for the
$^{244}$Pu
fundamental rotational band compared to the quonic rotor result (\textbf{B}%
), deformed algebra result from reference \cite{BonaRotor}
(\textbf{C}) and the usual quantum rigid rotor result
(\textbf{D}). Spectrum \textbf{B} was obtained using $q=0.99478$.

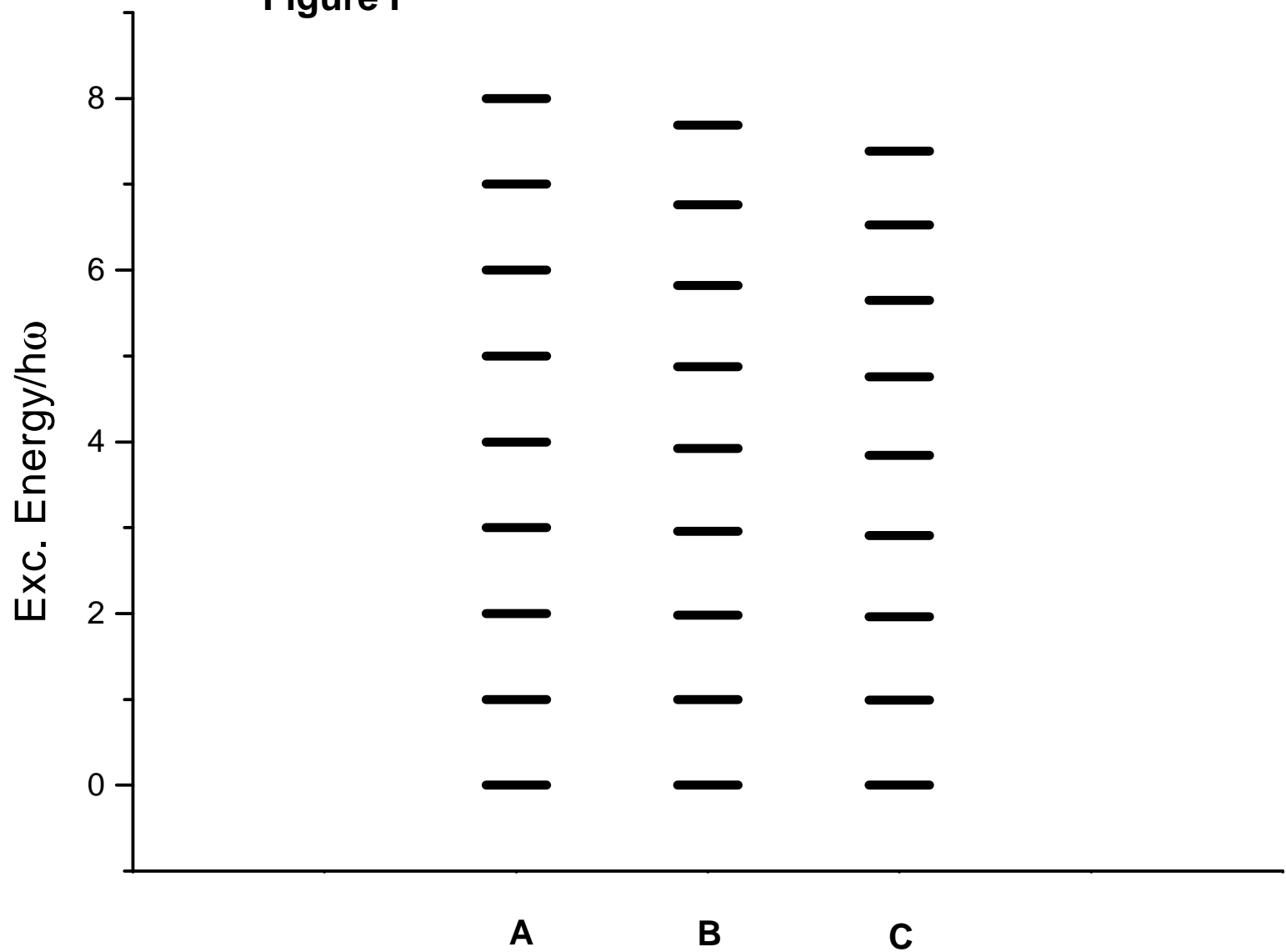
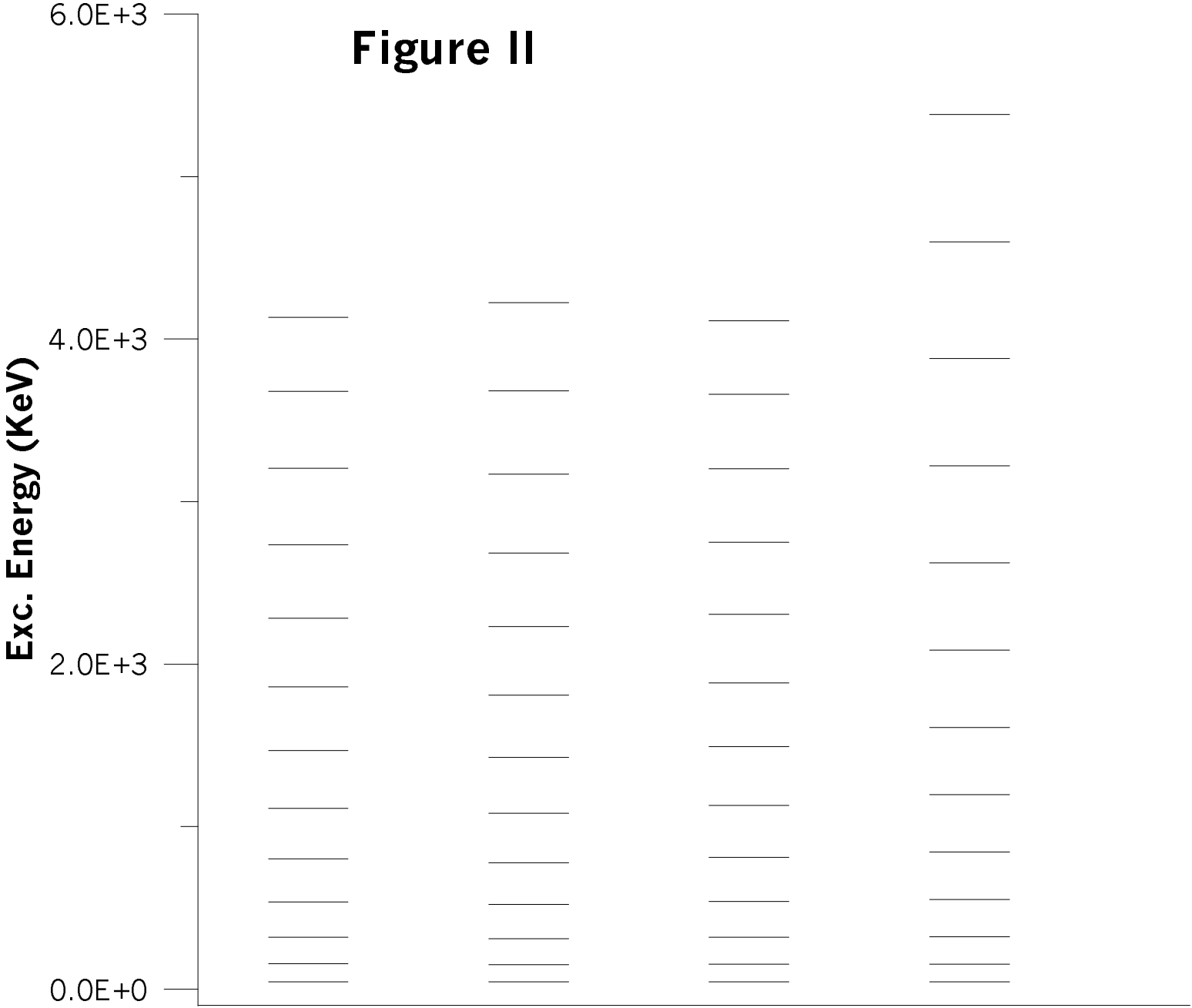
\end{document}